\documentclass{aa}
\usepackage{graphics,latexsym,amssymb,times,psfig}

\def\gsim{ \lower .75ex \hbox{$\sim$} \llap{\raise .27ex \hbox{$>$}} } 
\def\lsim{ \lower .75ex\hbox{$\sim$} \llap{\raise .27ex \hbox{$<$}} } 

\begin{document}

\title{The dividing line between FR~I and FR~II radio--galaxies }


\author{Gabriele Ghisellini \inst{1} and Annalisa Celotti \inst{2}}

\offprints{G. Ghisellini; gabriele@merate.mi.astro.it}
\institute{
Osservatorio Astronomico di Brera, via Bianchi 46, I--23807 Merate, Italy;
\and SISSA/ISAS, via Beirut 2-4, I--34014 Trieste, Italy
}

\date{Received 2001}
 
\titlerunning{FR~I--FR~II dividing line}
\authorrunning{G. Ghisellini \& A. Celotti}

\abstract{
In the radio -- host galaxy optical luminosity plane
FR~I and FR~II radio--galaxies are clearly divided.
Since the optical luminosity of an elliptical galaxy is an indication 
of the mass of its central black hole, we propose that the 
FR~I--FR~II dividing luminosity is a function of the mass of the
black hole powering the active nucleus.  
Furthermore, as the radio power gives an estimate of the total kinetic power 
carried by the jet, the FR~I--FR~II separation can be re--interpreted as 
occurring 
at a constant ratio between the jet power and the black hole mass.
There is also convincing evidence of a correlation between the radio
power and the luminosity in narrow emission lines. As the latter
results from photoionization by the radiation produced by accretion, we
can estimate the ionizing luminosity and find that the separation
luminosity can be also re--expressed as a constant accretion rate 
between $\sim 10^{-2}$--$10^{-3}$ of the Eddington one. 
This possibly regulates the accretion mode and the consequent
presence and characteristics of nuclear outflows.
\keywords{Galaxies: jets --- Galaxies: nuclei --- Radio continuum: galaxies}
}
\maketitle

\section{Introduction}

Among the strongest phenomenological clues on radio sources origin and
physics is the recognition by Fanaroff \& Riley (1974) that the
majority of 
radio galaxies can be classified into two morphological types
(FR~I and FR~II) according to where most of the luminosity is
radiated, i.e. edge darkened and edge brightened sources, and that
this division rather neatly translates into a separation in radio
power (respectively below and above $L_{178} \simeq 2.5\times 10^{33}
h_{50}^{-2}$ erg s$^{-1}$ Hz$^{-1}$ at 178 MHz).  
This division has
become even clearer and sharper when it has been found by Ledlow \&
Owen (1994, 1996) to be a function of the optical luminosity of the
host galaxy, in the sense of increasing dividing radio luminosity with
increasing optical luminosity of the host.

These pieces of evidence have prompted several physical 
interpretations, which
invoke either or both the interaction of the jet with the ambient medium
or/and nuclear intrinsic properties of the accretion and jet
formation processes. 
Among the former models the duality has been attributed to
the dynamics of a slowing jet in the ambient gas pressure (either the
whole jet or only the hot spot advance, Bicknell 1995; Gopal--Krishna
\& Wiita 1988, 2001), while the latter ones include the possible
different content of the jet plasma (electron--positron pairs
or normal electron--proton plasma), or the black hole spin 
(Reynolds et al. 1996; Baum et al. 1995; Meier 1999).

However there is a further ingredient which can be added to this
picture, namely the possibility of associating an estimate of the
central black hole mass to both the luminosity of the bulge component
in the host galaxies, as proposed by Kormendy \& Richstone
(1995) and by Magorrian et al. (1998), and the galaxy stellar velocity 
dispersion following the work by Ferrarese \& Merrit (2000) and 
Gebhardt et al. (2000).  
This information is a powerful new tool for tackling the long standing problem
of the black hole/galaxy formation, and also provides us with elements to
estimate the combination of accretion rate and radiative efficiency of
the nucleus of the active galaxies.

Interesting results have been already found in this context in
connection with the radio quite vs radio loud (possible) dichotomy,
where the latter objects appear to be associated with higher mass
black holes 
when objects of the two classes are chosen to have similar 
optical nuclear (AGN) luminosity (McLure \& Dunlop 2001). 


Here, we focus on the issue of the dichotomy between FR~I vs FR~II
radio--galaxies in the radio power -- host galaxy magnitude plane,
taking advantage of the new information on the black hole mass and the
indications of connections between the observed radio luminosity and
the intrinsic jet power and the luminosity dissipated in the accreting
matter flow. In other words through these correlations we translate
the separation between FR~I and FR~II 
into a critical value of the mass accretion rate.

The key steps (and assumptions) of our derivation are the following.
(i) The conversion between host optical magnitude and black hole mass; (ii)
the association of the radio luminosity to the jet kinetic power;
(iii) the connection of the radio luminosity with the optical
luminosity responsible for the photoionization of the [OII] narrow
emission lines.  The details and results for each of these three steps are the
content of the next section. In Section 3 we discuss our findings and in
Section 4  we present our conclusions.

\section{The FR~I--FR~II dividing line}

\subsection{Host optical luminosity and black hole masses}

For the conversion of host galaxy optical magnitude into central black hole
mass $M_{\rm BH}$ we adopt the relation presented in McLure \& Dunlop (2001).  
Specifically this is expressed in terms of the absolute optical
R--band magnitude of the host galaxy $M_{\rm R}$ as

\begin{equation}
\log(M_{\rm BH}/M_\odot) \, =\, -0.62(\pm0.08) \, M_{\rm R} -5.41(\pm 1.75).
\end{equation} 

By applying this correlation the range of absolute magnitudes of
Fig.~1 can be immediately re--expressed as a range of black hole
masses. We report in Fig.~1 the original plot presented by Ledlow \&
Owen (1996) with the mass reported on the upper x--axis.  The dividing
radio power between FR~I and FR~II results to be a linear (within the
errors) function of the black hole mass.
In other words for any given radio luminosity an FR~I morphology tends
to be systematically associated with the more massive black holes. 

In the following we use the connection between the jet radio power and
intrinsic nuclear luminosities. 
For an assumed efficiency this will allow us to re-express the radio
luminosity vs host galaxy magnitude plane in terms of mass accretion
rate vs black hole mass.


\subsection{Relation between radio luminosity and jet power}

Let us then start considering the relation between radio power and 
kinetic power output of the jet.  
It has been found that the radio
luminosity gives an estimate of the average power transported by the
jet to the outer lobes. 
In particular, several authors (Rawlings \& Saunders 1991; 
Rawlings 1992; Willott et al. 1999)
have found significant correlations between the radio luminosity
(and/or the luminosity in narrow lines) and the jet kinetic 
power $L_{\rm jet}$. 
Here we have adopted the correlation reported by Willott et al. (1999),
namely:

\begin{equation}
L_{\rm jet} \, =\, 3 \times 10^{21} L^{6/7}_{\rm 151}\,\,\,\,  
{\rm erg \, s^{-1}}, 
\end{equation}

\noindent
where $L_{151}$ [erg s$^{-1}$ Hz$^{-1}$ sr$^{-1}$]
is the monochromatic radio power at 151 MHz.
To convert this into a luminosity at 1.4 GHz (in W Hz$^{-1}$)
we have assumed a radio spectral index 
$\alpha=0.8$ [$L(\nu)\propto \nu^{-\alpha}$]
\footnote{We have also taken into account the different value of $H_0$ 
used in Willott et al. (1999) and in Ledlow \& Owen (1996).}.
We can thus determine the relation between $L_{\rm jet}$ and $M_{\rm BH}$. 
In Fig.~1 we show the resulting $L_{\rm jet}$ on the right hand side y--axis.  
It becomes apparent that the division between FR~I and FR~II corresponds
to a separation at $\sim$ {\it constant} $L_{\rm jet}/M_{\rm BH}$. 
Quantitatively this can be expressed as 

\begin{equation} 
{L_{\rm jet} \simeq  0.015  L_{\rm Edd}}, 
\end{equation}

\noindent
where $L_{\rm Edd}=1.3\times 10^{38} (M_{\rm BH}/M_\odot)$ erg s$^{-1}$
is the Eddington luminosity. 

\subsection{Relation between radio luminosity and accretion luminosity}

Finally, let us estimate  the nuclear radiative output, by 
considering the well established relation between the luminosity in
narrow emission lines, believed to result from photoionization by
the nuclear (accreting) radiation $L_{\rm ion}$, and the radio power. 
This appears to be particularly significant in the case of the
[OII] emission - while part of the [OIII] luminosity might be affected
by obscuration  (Baum \& Heckman 1989, Browne \& Jackson 1992, but see also 
Jackson \& Rawlings 1997). 
This relation has been presented by several
authors (Saunders et al. 1989, Rawlings 1992, Willott et al. 1999).
We consider here again the results by Willott et al. (1999) and adopt the
relation 

\begin{equation} 
L_{\rm ion} \, \sim \, 5\times 10^3\,  L_{151}.
\end{equation} 
\noindent
Through that, we simply convert the radio luminosity 
into an estimate of $L_{\rm ion}$ as shown in Fig.~2. 
The division between FR~I and FR~II then turns 
out to be a separation at {\it constant} $L_{\rm ion}/M_{\rm BH}$ and more
precisely is described by the relation 

\begin{equation} 
L_{\rm ion} \, \sim \, 6\times 10^{-3} L_{\rm Edd}.
\end{equation} 

\begin{figure}
\psfig{figure=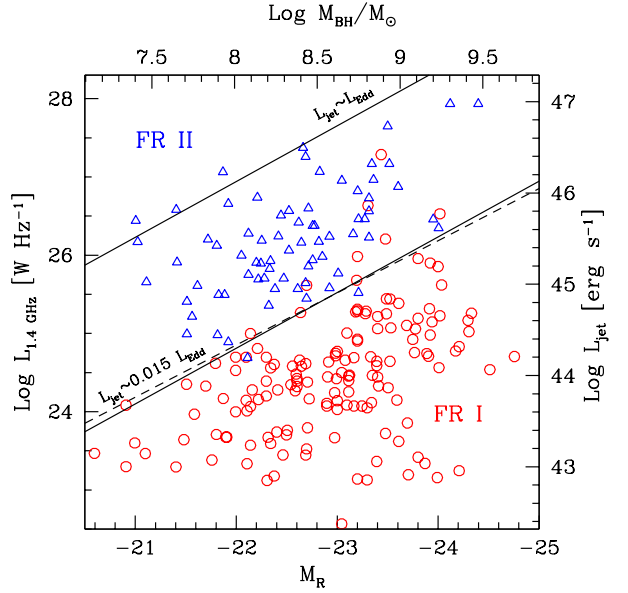,angle=0,width=9cm}
\vskip -0.5 true cm
\caption{The radio jet power -- host optical magnitude plane with the
line dividing FR~I from FR~II (dashed line, from Ledlow \& Owen 1996).
Triangles: FR~II; circles: FR~I. 
The two axis have been re--expressed as jet power vs black hole mass 
(right and upper axis). 
The two diagonal solid lines represent $L_{\rm jet} = 0.015 L_{Edd}$ and  
$L_{\rm jet} = L_{\rm Edd}$.}
\end{figure}

\begin{figure}
\psfig{figure=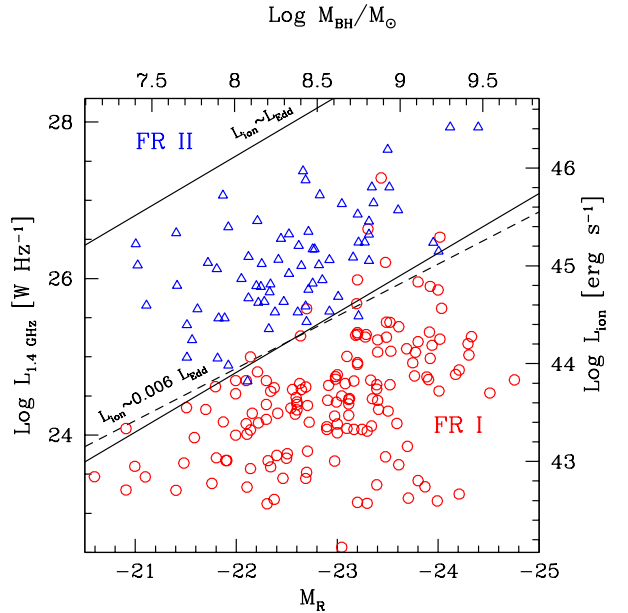,angle=0,width=9cm}
\vskip -0.5 true cm
\caption{The radio jet power -- host optical magnitude plane with the line 
sharply dividing FR~I from FR~II (dashed line, from  Ledlow \& Owen 1996). 
According to reasonably well established correlations this plane 
is equivalent to an accretion  power vs black hole  mass plane 
(right and upper axis). 
The central diagonal line represents 
$L_{\rm ion} \sim 6\times 10^{-3} L_{\rm Edd}$.}
\end{figure}

\section{Discussion}

We find that the separation in FR~I/FR~II morphology and
power appears to occur at a certain value of the luminosity 
over black hole mass ratio.
Although we neglected the significant dispersions of 
the assumed correlations, note that these would imply a fuzzier 
separation between the two classes, but would not alter the 
average behaviour we consider here. 
In the following we speculate about possible interpretations of this  
finding. 

\subsection{Are we finding a critical value of $\dot m$ where the 
accretion mode changes?}

It is tantalizing to speculate that the primary reason of the
FR~I--FR~II dichotomy lies in the different nature of their accretion
disks.  The found value of $L_{\rm ion}/L_{\rm Edd}\sim 6\times
10^{-3}$ suggests a critical value of $\dot m$, the accretion rate in
Eddington units, for which the mode of accretion changes
(within the uncertainties of the above correlations, e.g. Willot et al. 1999). 
This change might correspond to the
transition from a standard optically thick geometrically thin
efficient Shakura--Sunyaev (1973) disk to a radiatively
inefficient optically thin flow as an ion supported torus (Rees et
al. 1982) in the form of e.g. an advection dominated accretion flow
(ADAF, see e.g. Narayan, Garcia \& McClintock 1997), adiabatic
inflow--outflow (ADIOS, Blandford \& Begelman 1999) or a
convection dominated flow (CDAF, Narayan, Igumenshchev \& Abramowicz
2000).

If $L_{\rm ion}$ originates by the dissipation of 
the accretion power, 
$L_{\rm ion}\sim L_{\rm disk}=\eta\dot M_{\rm acc}c^2$
and if the efficiency $\eta$ is constant, at least within 
the FR~II population,
we  have that the FR~I--FR~II division line is quantitatively
described by

\begin{equation} 
\dot m \, \equiv \, {\dot M_{\rm acc} \over  \dot M_{\rm Edd} } \, \sim \, 
6\times 10^{-2} \eta_{-1}^{-1},
\end{equation} 

\noindent
where $\dot M_{\rm Edd}\equiv L_{\rm Edd}/c^2$
and $\eta = 0.1 \eta_{-1}$.  
Note that, within the sample considered by Ledlow \& Owen (1996), the
radio galaxies span a wide range of $\dot m$, between $\dot m \sim
10^{-4}$ and $\sim 10$, and nevertheless such transition has to be
rather sharp in order to produce such a well defined dividing line.


But how could be the accretion mode affect the large scale structure
of the radio galaxies? 

Observationally, we know that the structures of the parsec scale jet
of FR~I and FR~II are very similar and no difference in their
velocities appears to be present at these scales
(e.g. Giovannini et al. 2001).  On the other hand, it is believed that
on the kpc scale FR~I jets have velocities smaller than FR~II jets
(e.g. Begelman 1982, Bicknell 1984, Laing 1993): mildly relativistic
transonic jets could be more subject to Kelvin--Helmoltz 
instabilities, leading to the typical FR~I morphology (it is
conceivable that the deceleration is first due to the interaction with
circum--jet material).  However it is not clear at what scale an FR~I
jet decelerates.
Indeed, high values of the bulk Lorentz factor $\Gamma$ ($\sim$
10--15) are required to account for the spectral energy distribution
of high energy peak BL Lac objects (HBL) which are believed to be FR~I
whose jet is aligned with the line of sight (Ghisellini et al., 1998;
Tavecchio et al., 1998).  But the very same objects do not show the
extreme superluminal motion seen in the more powerful blazars thought
to be FR~II seen end--on (see e.g. Marscher 1999, Jorstad et
al. 2001).  It is thus possible that either deceleration occurs
between say a fraction of a parsec, where most of the emission is
produced, and the VLBI parsec scale, or that HBL are preferentially
seen at angles smaller than $1/\Gamma$ (resulting in a lower apparent
superlumimal velocity).  Independently of the cause, in this scenario
FR~I jets start highly relativistic and decelerate between the subpc
and the kpc scales.  
Indeed this is a crucial
ingredient in the model proposed by Bicknell (1995), who points to the
environment and the consequent deceleration as the main cause of the
FR~I--FR~II dichotomy.  

In this context, our findings suggest that it might be also the
accretion process itself to play a key role in the deceleration and
dichotomic behaviour, by affecting the pc--kpc scale environment.
Although at this point it might be premature to single out a
consistent model linking the accretion mode and the jet behaviour on
the pc--kpc scale, one could speculatively attribute it to the
presence of a wind -- produced by the disk itself and interacting with
and slowing down the relativistic jet -- becoming more important for
lower accretion rates, as predicted by some accretion scenarios (see
above and also numerical simulations by Stone, Pringle \& Begelman
1999). 
We can only speculate about the expected signatures of the
interaction of a wind and a relativistic jet.  This might lead
to the formation of shocks (similar to the analog {\it external
shocks} in gamma-ray bursts) and the efficient conversion of the bulk
kinetic energy into radiation (see e.g. Dermer 1999).  FR~I radio
galaxies -- and their aligned counterparts BL Lacs --  could
therefore be more efficient radiators than FR~II radio galaxies and
emission line blazars (although with a smaller absolute emitted
power).

Since for a given mass the accretion luminosity in low radiative
efficiency flows is expected to increase with $\dot M_{\rm acc}^2$
(see e.g. Narayan, Garcia \& McClintock, 1997) 
this scenario can naturally account 
for the lack of intense broad emission lines in FR~I sources 
and BL Lac objects (but allowing some sources, as BL Lac itself,
to show broad emission lines, albeit weak).
Note that in this case the absence of broad lines would not 
be ascribed to obscuration
(which remains a possibility in FR~II sources), but to the weak level
of the ionizing continuum, as suggested by the detection of 
non--thermal nuclei in HST images of FR~I sources
(Chiaberge, Capetti \& Celotti 2000).


\section{Conclusions}

We have shown that the FR~I--FR~II dividing line in the radio luminosity
vs optical host galaxy luminosity can be re--expressed as a line
of constant ratio between the jet and/or the disk accretion power and the 
Eddington luminosity.
This suggests that the FR~I--FR~II dichotomy could be controlled by 
the properties of the underlying accretion process more than 
(or in addition to) a different environment.

The specific value of the division, $L_{\rm ion}\sim 6\times 10^{-3}
L_{\rm Edd}$, could correspond to a change in the accretion
mode.
 
Since FR~I have, on average, larger black hole masses, they might 
be older or have accreted at a greater rate in the past 
(through e.g. mergings), and therefore it
is conceivable to argue that at least 
a fraction of them were FR~II radio--galaxies in the past.
This might account for the different evolution properties
of the two classes (see e.g. Urry \& Padovani 1995).

\begin{acknowledgements}
We thank S. Campana and G. Giovannini for discussions.
AC thanks the Italian MURST and ASI for financial support.
\end{acknowledgements}

\end{document}